\documentclass[prl,twocolumn,aps,superscriptaddress]{revtex4-1}
\usepackage{color}
\usepackage{amsmath}
\usepackage{graphicx}
\usepackage{dcolumn}
\usepackage{bm}
\usepackage{subfigure}

\begin{document}

\title{Double Dirac Semimetals in Three Dimensions}

\author{Benjamin J. Wieder} 
\affiliation{ Department of Physics and Astronomy, University of Pennsylvania, Philadelphia, PA 19104-6323, USA}
\author{Youngkuk Kim} 
\author{A. M. Rappe}
\affiliation{The Makineni Theoretical Laboratories, Department of Chemistry, University of Pennsylvania, Philadelphia, PA 19104-6323, USA}
\author{C. L. Kane}
\affiliation{ Department of Physics and Astronomy, University of Pennsylvania, Philadelphia, PA 19104-6323, USA}

\begin{abstract}
We study a class of Dirac semimetals that feature an eightfold-degenerate double Dirac point.   We show that 7 of the 230 space groups can host such Dirac points and argue that they all generically display linear dispersion.   We introduce an explicit tight-binding model for space groups 130 and 135, showing that 135 can host an intrinsic double Dirac semimetal -- one with no additional degeneracies at the Fermi energy.  We consider symmetry-lowering perturbations and show that uniaxial compressive strain in different directions leads to topologically distinct insulating phases.   In addition, the double Dirac semimetal can accommodate topological line defects that bind helical modes.   Potential materials realizations are discussed.
\end{abstract}

\maketitle

A striking consequence of symmetry and topology in the electronic structure of materials is the existence of protected degeneracies that guarantee semimetallic behavior.  Such degeneracies occur in graphene~\cite{graphene} (in the absence of spin-orbit interactions) as well as at the surface of a topological insulator (TI)~\cite{tireview}.  In 2011, Wan et al.~\cite{wan} pointed out that twofold degenerate Weyl points could occur in bulk 3 dimensional (3D) materials.   Such Weyl points are topologically protected, though they are ``symmetry prevented" in that they require broken inversion or time-reversal (T) symmetry to exist.   Crystal symmetries can lead to a richer variety of nodal semimetals.   Dirac semimetals~\cite{young,wang1,wang2}, which feature fourfold degenerate Dirac points protected by crystal symmetry, occur in two varieties.   Topological Dirac semimetals, such as Cd$_2$As$_3$ and Na$_2$Bi~\cite{liu1,liu2,borisenko}, exhibit Dirac points on a rotational symmetry axis due to a band inversion.  Nonsymmorphic Dirac semimetals, predicted in BiO$_2$~\cite{young} and  in BiZnSiO$_4$~\cite{steinberg}, conversely have Dirac points at high-symmetry points which are guaranteed by the underlying Space Group (SG) symmetry.  Additional classes of nodal semimetals include line nodes~\cite{wengliang,volovik,kee1,kee2,bernevig,kimkane,yu,xie,zwang} in 3D and Dirac semimetals in 2D~\cite{youngkane,schoop}.

In this paper we introduce and analyze a double Dirac semimetal that exhibits a single eightfold degeneracy point at a Brillouin Zone (BZ) corner.   We show that 7 of the 230 SGs host double Dirac points (DDPs) and argue that all of them generically have linear dispersion.   For two of the SGs (130 and 135) a DDP is guaranteed whenever the band filling is an odd multiple of four, while for the remaining five SGs the presence of DDPs depends on the band ordering.   We introduce an explicit tight-binding model for SGs 130 and 135 that demonstrates the DDP, and we study its low energy structure in detail.   Like the single (nonsymmorphic) Dirac semimetal, the double Dirac semimetal can be gapped into a trivial or topological insulator by applying strain.  In double Dirac semimetals, both phases can be achieved with compressive strain oriented along two different directions.   Moreover, in the double Dirac semimetal, spatially modulating the symmetry-breaking energy gap can lead to topological line defects that bind 1D helical modes.   Materials hosting DDPs are discussed at the end of the paper.

The existence of symmetry-protected degeneracies at a point ${\bf K}$ in the BZ can be ascertained by determining the dimension of the appropriate double-valued projective representations of the little group of ${\bf K}$.  This information has been tabulated for all 230 SGs~\cite{bradley}.  Table \ref{table1} lists all of the SGs with symmetry points that host Four-Dimensional Irreducible Representations (4DIR) that are also doubled by T symmetry.

\begin{table}
\begin{tabular}{l|ll|l|c|c}
\hline
\multicolumn{3}{c|}{Space Group} & ${\bf K}$ & Reps at ${\bf K}$ & Vector Reps\\
\hline
130 & $P4/ncc$ & $\Gamma_q D_{4h}^8$ & $A$ & $\Gamma_5^{\oplus 2}$(8) & $4E_u \oplus 3 A_{2u}$ \\
135 & $P4_2/mbc$ & $\Gamma_q D_{4h}^{13}$ & $A$ & $\Gamma_5^{\oplus 2}$(8) & $4E_u \oplus 3 A_{2u}$\\
218 & $P{\bar 4} 3 m$ & $\Gamma_c T_d^4$ & $R$ &  $\Gamma_6 \oplus\Gamma_7$(4),  $\Gamma_8^{\oplus 2}$(8)  & $2T_2$ \\
220 & $P{\bar 4}3d$ & $\Gamma_c^v T_d^6$ & $H$ &     $\Gamma_6 \oplus\Gamma_7$(4), $\Gamma_8^{\oplus 2}$(8) & $2T_2$ \\
222 & $Pn3n$ & $\Gamma_c O_h^2$ & $R$ &   $\Gamma_5$(4),  $\Gamma_6\oplus\Gamma_7$(8)& $3T_{1u}$ \\
223 & $Pm3n$ & $\Gamma_c O_h^3$ & $R$ & $\Gamma_5$(4),  $\Gamma_6\oplus\Gamma_7$(8) & $3T_{1u}$ \\
230 & $Ia3d$ & $\Gamma_c^v O_h^{10}$ & $H$ & $\Gamma_5$(4), $\Gamma_6\oplus\Gamma_7$(8) & $3T_{1u}$ \\
\hline
\end{tabular}
\caption{Space groups that host DDPs.   SGs are indicated in International notation as well as in Sh\"onflies notation, which indicates the crystal system and point group.  The momenta ${\bf K}$ are listed with symmetry labels for the 8DIRs, as well as for some 4DIRs.   The final column indicates the T-invariant vector representations of the point group contained in the tensor product $\Gamma^*\otimes\Gamma$ of the 8DIR at ${\bf K}$, indicating that in each case a linear dispersion is generic.}
\label{table1}
\end{table}

SGs 130 and 135 have the distinguishing feature that there is only a single 8DIR at the $A$ point.   Therefore groups of 8 bands ``stick together'', implying an insulator is only possible when the band filling is a multiple of 8.  Interestingly, Watanabe et al.~\cite{watanabe}(WPVZ) recently introduced a bound on the minimal filling for an insulator that applies to strongly-interacting systems.  The WPVZ bound for SG 130 is 8, in agreement with the band theory analysis, while for SG 135 the WPVZ bound of 4 disagrees with band theory~\cite{watanabe}.  Below we show that for SG 130, but not for SG 135, additional single Dirac points are present when the filling is an odd multiple of 4.  Since the energy of the single and DDPs will differ, SG 130 will generically host a semimetal with electron and hole pockets.   In contrast, in SG 135 the DDP is the only required degeneracy, so SG 135 can host an intrinsic double Dirac semimetal.  The fact that the symmetry-guaranteed  DDP is not covered by the WPVZ bound poses the interesting question of whether strong interactions can open a symmetry-preserving gap in SG 135.

For the remaining 5 SGs in Table \ref{table1} there are 4DIRs in addition to the 8DIRs at ${\bf K}$.   Therefore, the presence of DDPs at the Fermi level will depend on the band ordering.  This is similar to group IV semiconductors where band inversion at ${\bf k}=0$ leads to a fourfold degeneracy at $E_F$ in grey tin~\cite{greytin}.   In grey tin, the dispersion away from ${\bf k}=0$ is quadratic.   To determine whether the dispersion at the DDPs is linear we check whether the T-odd vector representation(s) are contained in the tensor product $\Gamma\otimes\Gamma^*$ of the 8DIR at ${\bf K}$ \cite{young}.   This is found by computing the character of the symmetric Kronecker square~\cite{bradley} of $\Gamma$ and using the orthogonality of characters to project onto the vector representation.   This analysis, which agrees with the specific example worked out below, predicts the multiple vector representations listed in Table \ref{table1}.  Therefore in all cases the dispersion near ${\bf K}$ will generically be linear.   The DDP is anisotropic for the tetragonal structures 130 and 135, while for the remaining cubic structures it is isotropic.

\begin{figure}
\includegraphics[width=8.5cm]{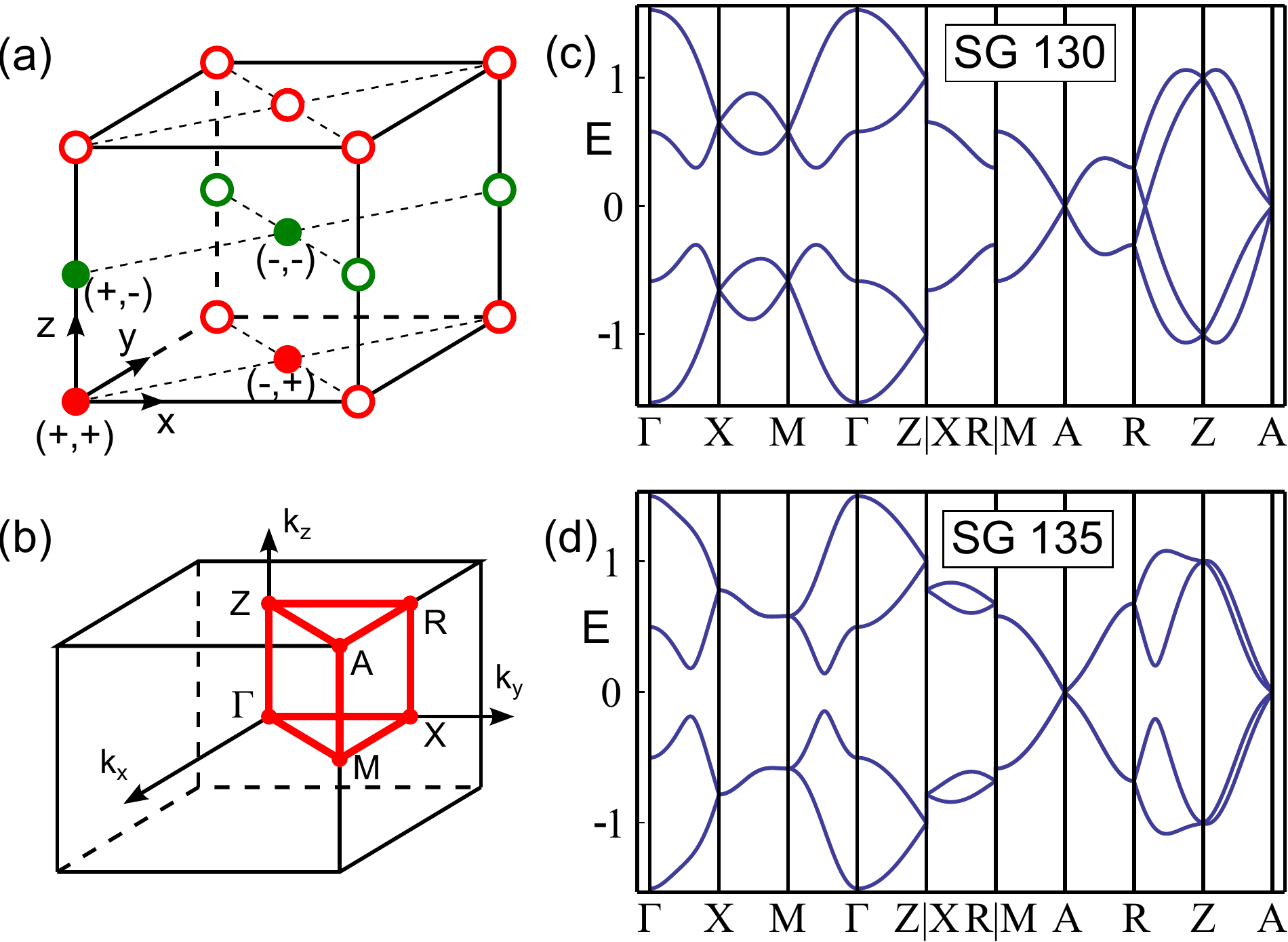}
\caption{(a) Model lattice for the common tetragonal structure of SGs 130 and 135.  The solid lines indicate the four-site unit cell, and solid circles denote the 4 sublattices labeled by $(\tau^z,\mu^z) = (\pm, \pm)$.   (b) Tetragonal Brillouin zone.   (c) Energy bands for SG 130, described by the tight-binding model (\ref{h0},\ref{v130}) with
$t_{xy}=1$, $t_z=0.5$ and $\lambda_1=\lambda_2=\lambda_3=0.3$.   In addition to the DDP at $A$, there is a symmetry-guaranteed Dirac point on the line $Z$-$R$.  (b)  Bands for SG 135, described by (\ref{h0},\ref{v135}) with
$t_{xy}=1$, $t_z=0.5$, $t_1'=t_2'=0.3$, $\lambda_1'=0.3$, $\lambda_2' = 0.1$, $\lambda_3'=0.25$.   There is a single DDP at $A$ with no other crossings.    }
\label{Fig1}
\end{figure}

We now introduce an explicit tight-binding model for SGs 130 and 135.  These SGs share the same  tetragonal structure and are characterized by the symmetry generators in Table \ref{table2}.  We introduce a unit cell (Fig. \ref{Fig1}(a)), with 4 sublattices indexed by $(\tau^z,\mu^z) = (\pm 1,\pm 1)$ associated with basis vectors ${\bf d} = \tfrac{1}{2}[(1-\tau^z) (\tfrac{1}{2}\tfrac{1}{2}0) + (1-\mu^z)(00\tfrac{1}{2})]$.  
This can be viewed as a distortion of a parent Bravais lattice~\cite{youngkane} in which the 4 sublattices are related by pure translations.   Nearest neighbor hopping on this parent lattice is described by
\begin{equation}
{\cal H}_0({\bf k}) = t_{xy}\tau^x\cos\tfrac{k_x}{2}\cos\tfrac{k_y}{2} + t_z \mu^x\cos\tfrac{k_z}{2},
\label{h0}
\end{equation}
where we choose a gauge in which the state associated with sublattice $(\tau^z,\mu^z)$ has phase
$\exp i {\bf k} \cdot {\bf d}$, so
\begin{equation}
{\cal H}({\bf k} + {\bf G}) = e^{-i {\bf G}\cdot {\bf d}(\tau^z,\mu^z)} {\cal H}({\bf k}) e^{i {\bf G}\cdot{\bf d}(\tau^z,\mu^z)}.
\label{gauge}
\end{equation}

\begin{table}
\begin{tabular}{c|c|c|c}
\multicolumn{2}{c|} {SG 130} & \multicolumn{2}{c} {SG 135} \\
\multicolumn{1}{c}{$\{g|{\bf t}\}$} & \multicolumn{1}{c|}{$D(\{g|{\bf t}\})$} & \multicolumn{1}{c}{$\{g|{\bf t}\}$} & \multicolumn{1}{c}{$D(\{g|{\bf t}\})$ } \\
\hline
$\{C_{4z}|000\}$ & $ e^{i\pi \sigma^z/4}$ & $\{C_{4z}|00\tfrac{1}{2}\}$& $ \mu^x e^{i\pi\sigma^z/4}$ \\ 
$\{C_{2x}|\tfrac{1}{2}\frac{1}{2}0\}$ & $i\tau^x\sigma^x$ & $\{C_{2x}|\tfrac{1}{2}\frac{1}{2}0\}$ & $i\tau^x\sigma^x$ \\
$\{I|\frac{1}{2}\frac{1}{2}\frac{1}{2}\} $&$\tau^x\mu^x$ & $\{I|000\}$ & $1 $\\
\end{tabular}
\caption{Symmetry generators for SGs 130 and 135, along with their representations in the sublattice-spin space.}
\label{table2}
\end{table}

SGs 130 and 135 involve lowering the translational symmetry while keeping different nonsymmorphic glide and screw symmetries.   The symmetry generators are represented by operators on the 8-dimensional spin and sublattice space $D(\{g|{\bf t}\})$ listed in Table \ref{table2}.   In addition, T symmetry is represented by $\Theta = i\sigma^y K$.  Symmetry-lowering perturbations ${\cal H} = {\cal H}_0 + V$ must satisfy
\begin{eqnarray}
V(g{\bf k}) &= &D(\{g|{\bf t}\})^\dagger V({\bf k}) D(\{g|{\bf t}\}), \\
V(-{\bf k}) &= &\Theta V({\bf k}) \Theta^{-1}.
\label{symmetries}
\end{eqnarray}
It is straightforward to enumerate the allowed terms for each SG.   In general, there are 28 terms consistent with inversion and T.   Eqs. \ref{gauge}-\ref{symmetries} determine the ${\bf k}$ dependence of each term.   Here we consider a subset of these terms that are sufficient to lift all nonessential degeneracies.
\begin{eqnarray}
V_{130} &= \lambda_1 \tau^z\mu^y \cos\tfrac{k_z}{2} + \lambda_2 \tau^z(\sigma^x \sin k_y - \sigma^y \sin k_x) \nonumber\\
 &+ \lambda_3 \tau^x\mu^z (\sigma^x\sin\tfrac{k_x}{2}\cos\frac{k_y}{2}+\sigma^y\cos\tfrac{k_x}{2}\sin\tfrac{k_y}{2}),
\label{v130}
\end{eqnarray}
and
\begin{eqnarray}
V_{135} &= t_1' \mu_z (\cos k_x - \cos k_y) +
t_2' \tau^y\mu^y \sin\frac{k_x}{2} \sin\frac{k_y}{2} \cos\frac{k_z}{2}  \nonumber \\
+ &\lambda_1' \tau^y\mu^x ( \sigma^x \sin\frac{k_x}{2}\cos\frac{k_y}{2}
+\sigma^y\cos\frac{k_x}{2}\cos\frac{k_y}{2}) \sin\frac{k_z}{2} \nonumber \\
+ &\lambda_2' \tau^x\mu^y ( \sigma^x \cos\frac{k_x}{2}\sin\frac{k_y}{2}+ \sigma^y\sin\frac{k_x}{2}\cos\frac{k_y}{2}) \sin\frac{k_z}{2} \nonumber \\
&+ \lambda_3' \tau^y\mu^z \sigma^z \cos\frac{k_x}{2}\cos\frac{k_y}{2}(\cos k_x-\cos k_y).
\label{v135}
\end{eqnarray}

Fig. \ref{Fig1}(c,d) show energy bands associated with these models.  Each band is at least doubly degenerate.   Both cases feature a DDP at $A$ with linear dispersion in all directions.   SG 130 features an additional fourfold crossing along the line $Z$-$R$.  
This crossing is protected by T, inversion and the $C_{2x}$ screw, whose axis is displaced from the center of inversion.   This guarantees that the Kramers-degenerate pairs of states on this line share the same eigenvalue of the $C_{2x}$ screw, allowing pairs with different eigenvalues to cross.   A similar crossing occurred in the Dirac ring found in Ref. \onlinecite{kee1} and was locally characterized in Ref. \onlinecite{fangfu}.   In fact, this crossing is {\it guaranteed} by symmetry, since it is not possible to eliminate it by reordering the bands at $Z$ or at $R$.   This pattern of degeneracies guarantees that groups of 8 bands stick together, independent of the DDP at $A$, and appears to be correlated with the WPVZ bound.   

Since the additional Dirac points need not be at the same energy as the DDP, SG 130 will generically be a semimetal with electron and hole pockets.   In contrast, SG 135 has no additional Dirac points, so it can host an intrinsic double Dirac semimetal.   However, we find that for $\lambda_2' > \lambda_1'$ there are additional single Dirac points along the lines $A$-$Z$.  These arise due to a ``velocity inversion" transition at $\lambda_2'=\lambda_1'$, which we analyze below.  A similar velocity inversion occurs in 130 for $\lambda_3 > 2\lambda_2$.

We now focus on SG 135 and consider the low energy structure near the DDP.   There are no symmetry-respecting terms at $A$ that lift the degeneracy.   To determine the terms linear in ${\bf k}$ we identify the T-odd operators transforming in the vector representations of the point group $D_{4h}$.  Using (\ref{gauge}-\ref{symmetries}), the representations of the symmetry operations at $A$ are $d_A(\{C_{4z}|00\frac{1}{2}\}) = \tau^z\mu^x \exp i\pi \sigma^z/4$, $d_A(\{C_{2x}|\frac{1}{2}\frac{1}{2}0\}) = \tau^y\mu^z\sigma^x$ and $d_A(\{I|000\}) = \mu^z$.  Also, $\Theta_A = i\mu^z\sigma^y K$.  We find 4 (3) terms with $E_u$ ($A_{2u}$) symmetry, in agreement with the general analysis of Table \ref{table1}.  The ${\bf k}\cdot {\bf p}$ Hamiltonian at $A$ is
\begin{eqnarray}
{\cal H_A}&= (u_0 \tau^x\mu^y\sigma^x + u_1 \tau^y\mu^x\sigma^y + u_2 \mu^y\sigma^y + u_3 \tau^z\mu^y\sigma^x) k_x \nonumber \\
       +& (u_0 \tau^x\mu^y\sigma^y + u_1 \tau^y\mu^x\sigma^x + u_2 \mu^y\sigma^x - u_3 \tau^z\mu^y\sigma^y) k_y \nonumber\\
       &+(v_1 \mu^x + v_2 \tau^y\mu^y + v_3 \tau^x\mu^y\sigma^z) k_z.
\end{eqnarray}
This leads to dispersion
\begin{eqnarray}
E^2_{\pm}&({\bf k})=(|{\bf u}|^2 + u_0^2)(k_x^2+k_y^2) + |{\bf v}|^2 k_z^2 \hskip .4in\nonumber\\
&\pm2\sqrt{ 4 k_x^2 k_y^2 |{\bf u}|^2 u_0^2 - (k_x^2+k_y^2) k_z^2 ({\bf u}\cdot {\bf v})^2},
\label{ha}
\end{eqnarray}
where ${\bf u}=(u_1,u_2,u_3)$ and ${\bf v}=(v_1,v_2,v_3)$.   When $|u_0| = |{\bf u}|$, one of the branches vanishes on the line $k_x=k_y$, $k_z=0$, identifying the velocity inversion transition along $A$-$Z$ discussed above.   From the tight-binding model, we have ${\bf u} = (\lambda_1',0,0)$, $u_0 = \lambda_2'$.   Therefore, we identify $|u_0| < |{\bf u}|$ with the intrinsic double Dirac semimetal phase with no additional degeneracies.   

\begin{table}
\begin{tabular}{l|l|l}
\hline
$A_{1g}$ & $1$    & Double Dirac SM \\
$A_{2g}$ & $\tau^x\mu^z$\quad $\tau^y\mu^z\sigma^z$\quad $\tau^z$ & Weak TI \\
$B_{1g}$ & $\mu^z$ &   Strong TI \\
$B_{2g}$ & $\tau^x$\quad $\tau^y\sigma^z$\quad $\tau^z\mu^z$ &  Weak TI \\
$E_g^{x,y}$ & $\tau^y\sigma^{x,-y}$ \quad$\tau^y\mu^z\sigma^{x,y}$ & Dirac Line/Point SM \\
\hline
$A_{1u}$ & $\tau^x\mu^y$ \quad $\tau^y\mu^y\sigma^z$ \quad $\mu^x\sigma^z$ & Strong TI \\
$A_{2u}$ & $\tau^z\mu^z\sigma^z$ & Dirac Point SM\\
$B_{1u}$ & $\mu^y\tau^z$ &Weyl Point SM\\
$B_{2u}$ & $\tau^x\mu^x\sigma^z$\quad $\tau^y\mu^x$\quad $\mu^y$ &Weyl Point SM\\
$E_u^{x,y}$ & $\tau^x\mu^x\sigma^{x,-y}$ \quad $\tau^z\mu^x\sigma^{x,y}$&Weyl Point  SM\\
&   $\tau^y\mu^y\sigma^{y,-x}$\quad$\mu^x\sigma^{y,-x}$\\
\hline
\end{tabular}
\caption{Perturbations to the DDP in SG 135, classified by their symmetry under the $D_{4h}$ point group.   The resulting insulating and semimetallic (SM) phases are indicated.}
\label{table3}
\end{table}

\begin{figure}
\includegraphics[width=7cm]{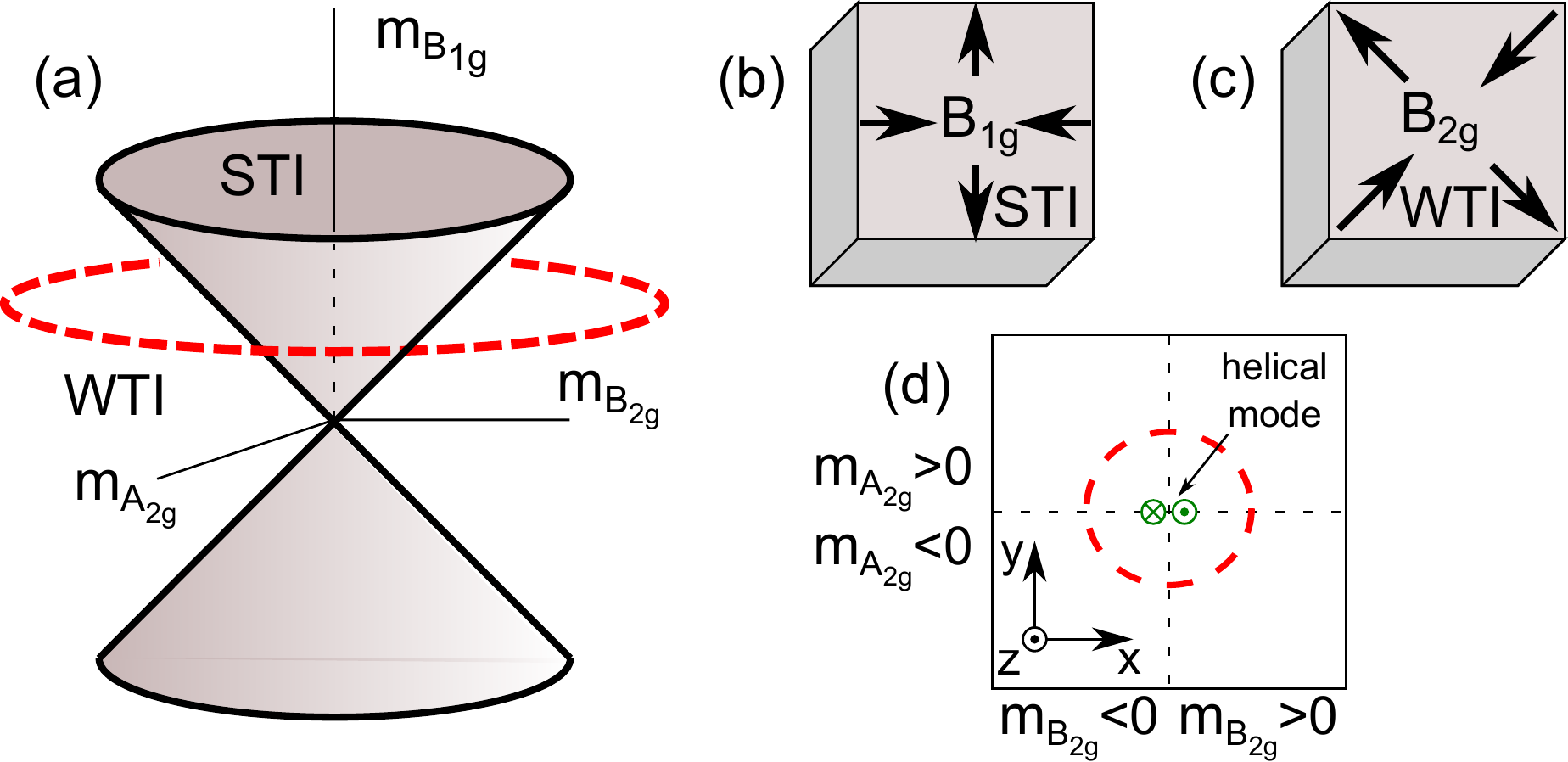}
\caption{(a) Phase diagram as a function of 3 symmetry-breaking perturbations.  A topologically nontrivial loop in the WTI phase is indicated by the dashed circle.  (b,c)  Uniaxial strain along different directions leads to topologically distinct insulating states.  (d)  A topological line defect in an insulating state binds a gapless 1D helical mode.}
\label{Fig2}
\end{figure}

We next consider the long-wavelength symmetry-breaking perturbations that open energy gaps and identify the resulting phases that arise.     
General symmetry-breaking perturbations are classified by their symmetries under the $D_{4h}$ point group, which can be determined from $d_A(\{g|{\bf t}\})$, as above.   The possible T-invariant perturbations at $A$ are listed in Table \ref{table3}.    There are many terms, and their effects depend on the form of the velocity terms.  In order to organize the behavior, we first fix the velocity terms and examine the terms that can open a gap in the spectrum.   We find that there are precisely four mass terms that arise due to perturbations with specific symmetries.   The Hamiltonian has the form
\begin{eqnarray}
{\cal H} &= u_2 \mu^y(\sigma^y k_x + \sigma^x k_y) + v_1 \mu^x k_z + m_{A_{2g}} \tau^x\mu^z + \nonumber\\
&m_{B_{2g}} \tau^z\mu^z + m_{B_{1g}} \mu^z + m_{A_{1u}} \tau^y\mu^y\sigma^z.
\label{masses}
\end{eqnarray}
The four mass terms are the unique terms from Table \ref{table3} that anticommute with all three of the velocity terms and open a gap.  They fall into two groups.   ${\bf m}_1 = (m_{A_{2g}},m_{B_{2g}})$ and ${\bf m_2} = (m_{B_{1g}},m_{A_{1u}})$ anticommute among themselves but commute with each other, leading to an energy gap 
\begin{equation}
E_{\rm gap} = 2\left||{\bf m}_1| - |{\bf m}_2|\right|.
\end{equation}   
${\bf m}_1$ and ${\bf m}_2$ thus define topologically distinct phases.   By evaluating the parity eigenvalues in the tight-binding model~\cite{fukane2}, we conclude that ${\bf m}_1$ defines a $(0;110)$ weak TI, while ${\bf m}_2$ defines a $(1;001)$ strong TI~\cite{fukane1}.   While these topological indices depend on the details of the band structure away from the DDP, the difference between them for the ${\bf m}_1$- and ${\bf m}_2$-dominated phases is robust.      To visualize the phases, Fig. \ref{Fig2}(a)  shows a 3D phase diagram for $m_{A_{1u}}=0$.    In this restricted space, the boundary between the STI and the WTI is a cone, with the STI (WTI) inside (outside) the cone.    In the more general 4D phase diagram, the WTI and STI phases appear symmetrically.

Different combinations of velocity terms lead to different choices for the anticommuting mass terms.  However, for any choice of velocity terms, there is always one anticommuting term (or linear combination of terms) with each of the symmetries in Eq. \ref{masses}.   This means that in the absence of fine tuning, a generic perturbation with a given symmetry induces the corresponding mass term and opens a gap.  The general structure of Fig. \ref{Fig2}(a) remains, except that when the inversion-symmetry-breaking term $m_{A_{1u}}$ is present the boundary between the STI and WTI phases broadens to include an intermediate Weyl semimetal phase~\cite{murakami}.

The dependence of the topological state on $m_{B_{1g}}$ and $m_{B_{2g}}$ provides a mechanism for controlling topological states using strain.  As indicated in Fig.\ref{Fig2}(b,c), uniaxial strain along the $x$ or $y$ directions induces a perturbation with a combination of $A_{1g}$ and $B_{1g}$ symmetry, while uniaxial strain along the diagonal $x\pm y$ directions has $A_{1g}$ and $B_{2g}$ symmetry.  Therefore these two kinds of compressive strain lead to topologically distinct insulators.   

DDPs can also be differentiated from single Dirac semimetals by the existence of two distinct anticommuting mass 
terms that lead to the same topological phase.   For a single Dirac semimetal with $4 \times 4$ Dirac matrices, the general structure of Clifford algebras predicts that there is only a single T-invariant mass term.
For $8 \times 8$ Dirac matrices there are {\it two} independent T-invariant 
mass terms.   
This means that the space of gapped states has a nontrivial first homotopy group, indicated by the dashed circle in Fig. \ref{Fig2}(a),
allowing topologically nontrivial {\it line defects} (Fig. \ref{Fig2}(d)).  Line defects in a 3D insulator in class AII have a $Z_2$ topological invariant characterizing the $3+1$D ${\cal H}({\bf k},\theta)$~\cite{teokane}.    When nontrivial, this guarantees that a 1D helical mode is bound to the line, similar to the helical mode bound to a lattice dislocation in a weak TI~\cite{ran}.   Without a lattice dislocation,  this $Z_2$ invariant is inaccessible in a 4 band system because it derives via dimensional reduction~\cite{qihugheszhang} from a {\it third} Chern number in $3+1+2$D, which requires at least 8 bands.     
To establish that a line defect binds a 1D helical mode, we follow the analysis of Ref. \onlinecite{teokane}, and consider a simple model with $m_{B_{2g}} = a x$, $m_{A_{2g}} = a y$ and $u_1 = v_2 = v$.  ${\cal H}^2$ in (\ref{masses}) then has the form of a harmonic oscillator, and there is a single pair of modes with $E=\pm v k_z$ localized near $x=y=0$.

We finally briefly consider perturbations in Table \ref{table3} with the remaining symmetries, which lead to Weyl or Dirac semimetals.   The $E_g$ perturbations lead to either Dirac points or a Dirac ring, with fourfold-degenerate crossings.   The inversion breaking-perturbations generally lead to a Weyl semimetal with twofold crossings, except for $A_{2u}$, where the remaining $C_{2z}$ and glide mirror symmetries guarantee doubly-degenerate states for $k_x=k_y=\pi$ with the same $C_{2z}$ eigenvalue.   This allows degenerate bands to cross along $A$-$M$, protecting a Dirac point even though inversion is violated.

\begin{figure}
\includegraphics[width=8cm]{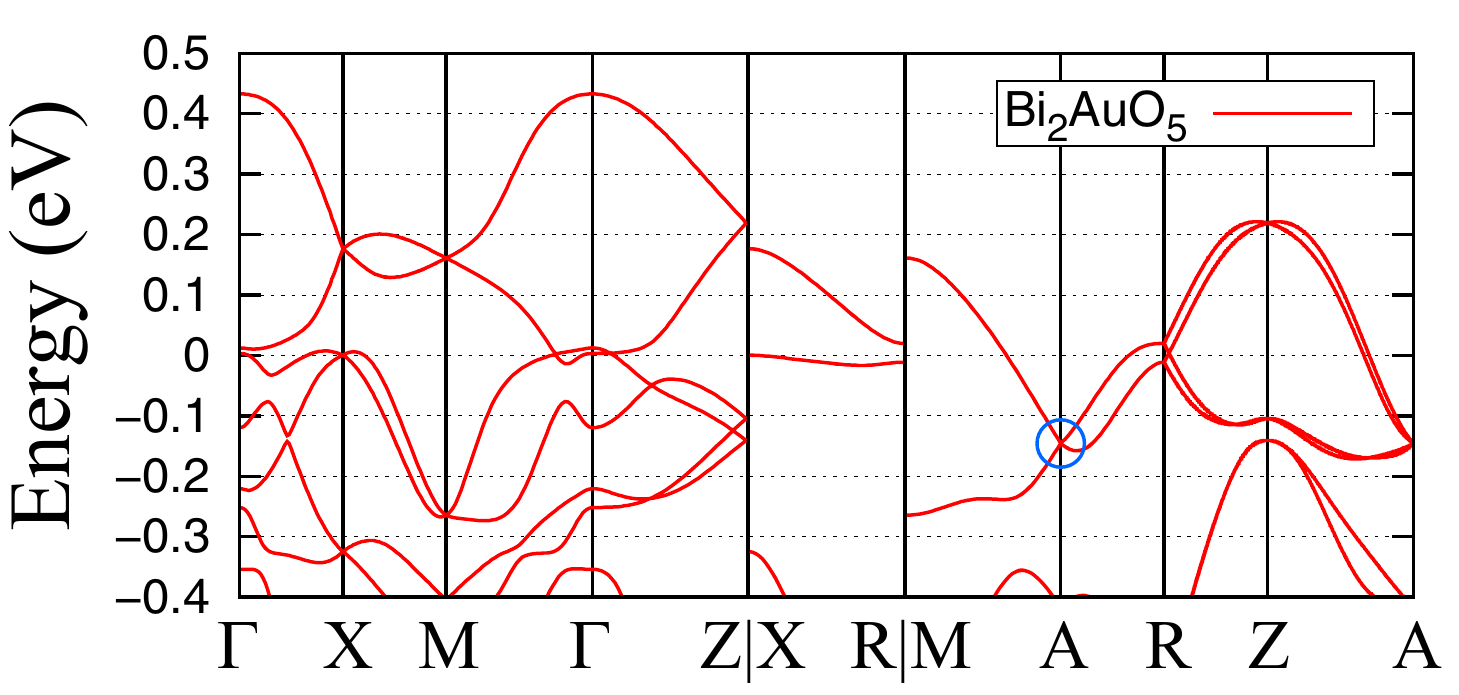}
\caption{(a) Band structure of Bi$_2$AuO$_5$ in SG 130, obtained from first-principles calculations. The DDP appears at $A$ with extra Dirac points along $R$-$Z$.  See the supplementary material for the methods.}
\label{Fig3}
\end{figure}

Encouragingly, the DDP appears to be feasible in known materials~\cite{appendixnote}. For example, a ternary bismuth aurate, Bi$_2$AuO$_5$ in SG 130, which has been synthesized in a single crystal~\cite{geb96p364}, hosts a DDP at $A$ close to the Fermi level, with additional four-fold degeneracies appearing on $Z$-$R$ (Fig. \ref{Fig3}).  As for materials in SG 135, the \textsc{Materials Project}~\cite{Jain2013} shows that a class of oxide materials isostructural with Pb$_3$O$_4$~\cite{Gavarri81p81}, including Sn(PbO$_2$)$_2$, Pb$_3$O$_4$, and Mg(PbO$_2$)$_2$, host the DDPs in the valence energy regime. Although they are semiconductors with electron filling 8, their atomic structure allows for a potential route towards a material design that shifts the Dirac point near the Fermi level. The number of atoms per each species in a unit cell is 4 (mod 8) and thus allows for filling 4 when suitably substituting atoms with an odd number of valence electrons.   
In SG 223, GaMo$_3$~\cite{Bornard73p205} hosts a DDP.   There is reason for optimism that with appropriate band structure engineering, an intrinsic double Dirac semimetal can be realized.

\begin{acknowledgments}
We thank Eugene Mele and Steve Young for helpful discussions.  This work was supported by 
NSF grant DMR 1120901 and a Simons Investigator grant to CLK from the Simons Foundation.
AMR acknowledges the support of the DOE, under grant DE-FG02-07ER46431.

\end{acknowledgments}

\renewcommand{\thefigure}{S\arabic{figure}}
\setcounter{figure}{0}
\renewcommand{\theequation}{S\arabic{equation}}
\setcounter{equation}{0}

\section{Supplementary Material for Double Dirac Semimetals in Three Dimensions}

\subsection{Computational Methods}

We perform \textit{ab initio} calculations based on Density Functional Theory (DFT) to find the band structures of Bi$_2$AuO$_5$, Sn(PbO$_2$)$_2$, Pb$_3$O$_4$, and GaMo$_3$. We use the Perdew--Burke--Ernzerhof--type generalized gradient approximation~\cite{Perdew96p3865} as found in the \textsc{Quantum Espresso} package~\cite{Giannozzi09p395502}.  Norm--conserving, optimized, designed nonlocal pseudopotentials are generated using the \textsc{Opium} package~\cite{Rappe90p1227, Ramer99p12471}.  An energy cutoff of 680 eV is used for the plane--wave basis. A fully-relativistic non--collinear scheme is employed to describe spin-orbit interaction.  The atomic structures are obtained from the inorganic crystal structure database (ICSD) database~\cite{ICSD}. 

\subsection{Double Dirac Points in Real Materials}
Here we present our DFT band structures of real materials that host the double Dirac point. 
We have searched the \textsc{Materials Project}~\cite{Jain2013} to find stable crystals in the space groups 130, 135, 218, 220, 222, 223, and 230, resulting in the following set of candidate materials.

\subsubsection{Bi$_2$AuO$_5$ in space group 130}

\begin{figure}[tH]
\includegraphics[width=0.45\textwidth]{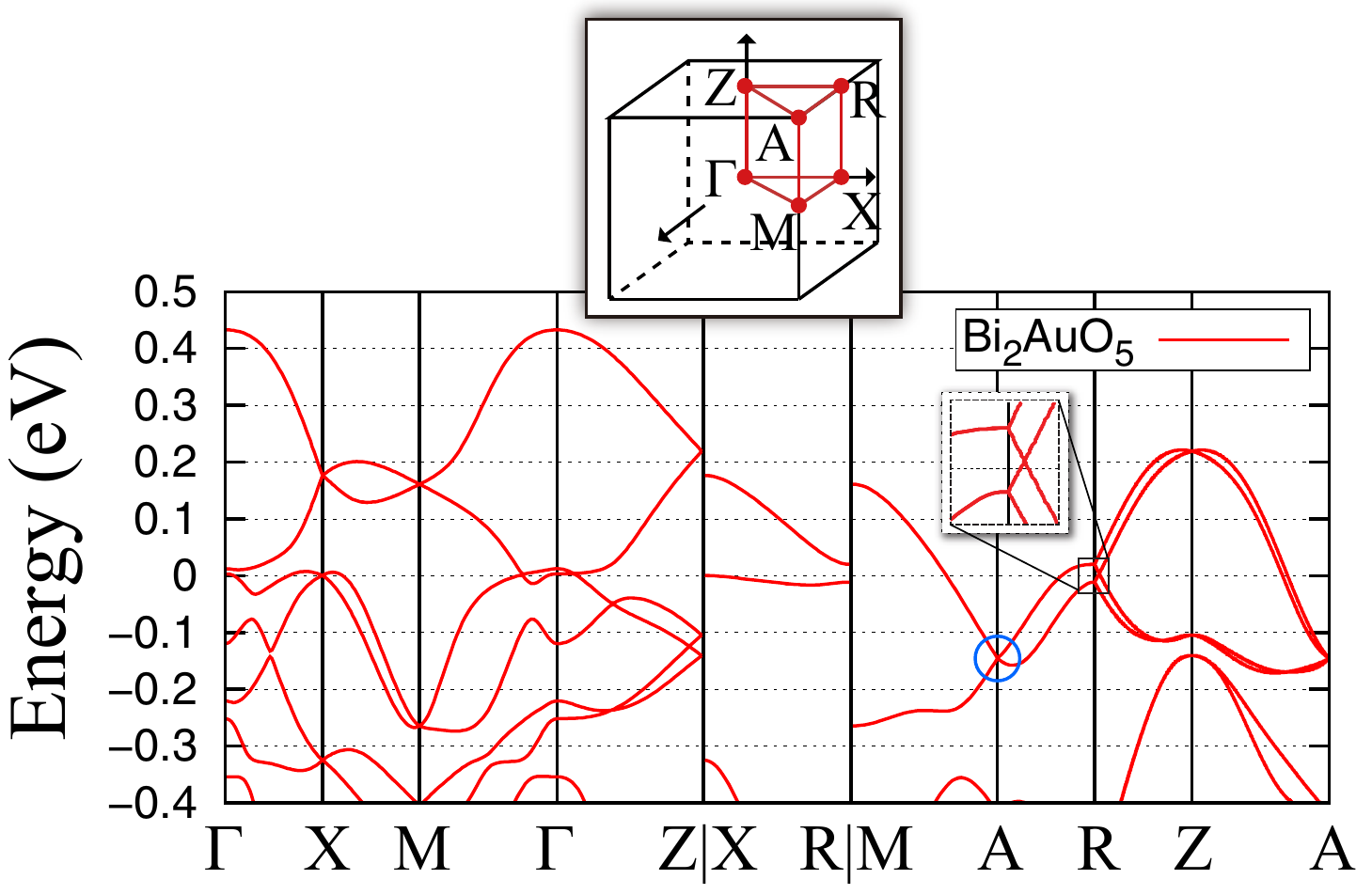}
\caption{\label{fig1}
Electronic band structure along high--symmetry lines in Bi$_2$AuO$_5$ in space group 130.
} 
\end{figure}

The energy band structure of Bi$_2$AuO$_5$ in space group 130 is shown in Fig.\ \ref{fig1}. The band structure is drawn along the high--symmetry lines of the tetragonal BZ, shown in the upper panel. The Fermi level is set to $0$ eV. The double Dirac point (indicated by a blue circle) appears at $A$ in good agreement with the results of the four-site tight-binding model. There is an additional Dirac point in $R$-$Z$, guaranteed by symmetry, which is magnified in the upper inset.  We expect that Bi$_2$AuO$_5$, which has been synthesized in a single-crystal phase~\cite{geb96p364}, should realize a double Dirac semimetal.

\subsubsection{Sn(PbO$_2$)$_2$ in space group 135}

\begin{figure}[h]
\includegraphics[width=0.45\textwidth]{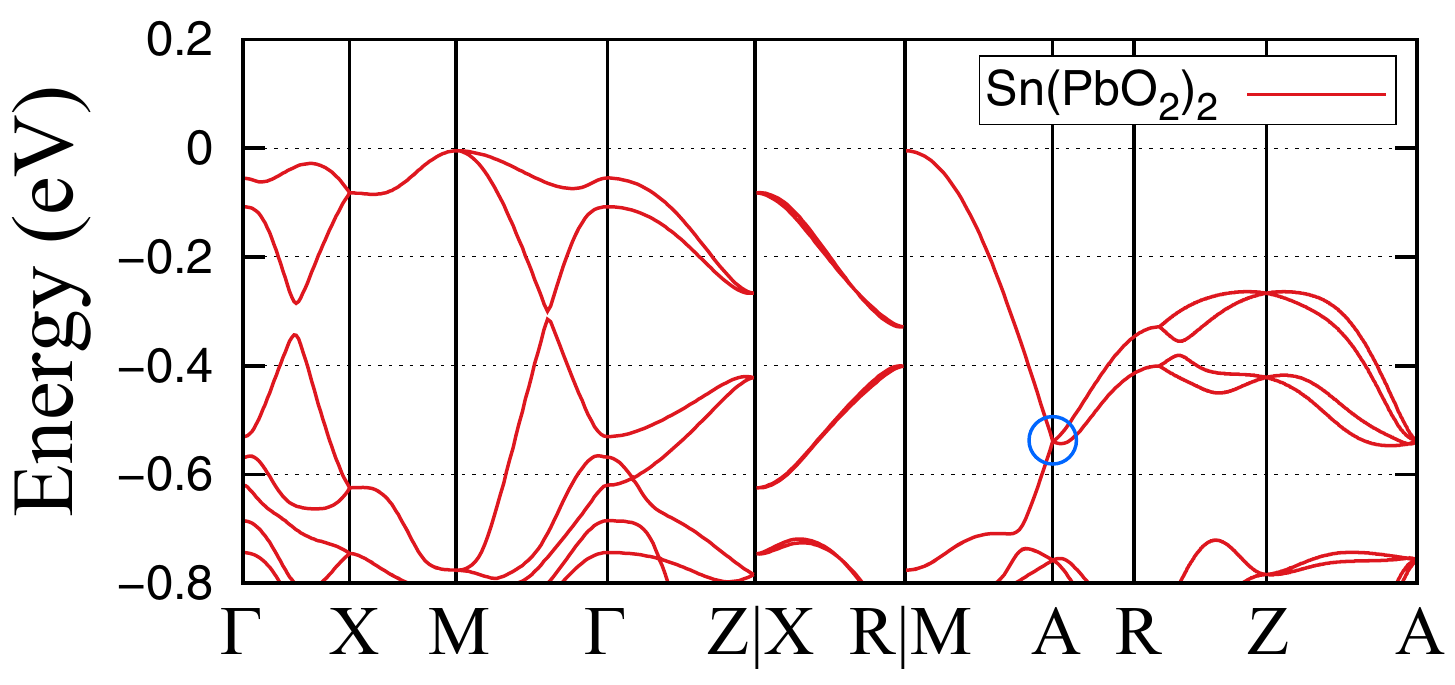}
\caption{\label{fig2}
Electronic band structure of Sn(PbO$_2$)$_2$ in space group 135. 
}
\end{figure}

In Fig.\ \ref{fig2} , the electronic energy band structure of Sn(PbO$_2$)$_2$ in space group 135 is presented.  The Fermi level is set to $0$ eV and  the valence bands are plotted along high--symmetry lines of the tetragonal BZ.  The double Dirac point is enclosed by a blue-colored circle. The number of Sn, Pb, and O atoms in the unit cell are 4, 8, and 16, respectively, and the number of valence electrons of Sn, Pb, and O are 14, 14, and 6, respectively, and thus the total number of valence electrons is 72. Although Sn(PbO$_2$)$_2$ is an insulator with filling zero (mod 8), a double Dirac point appears in the valence band at 0.54 eV below the Fermi level. The atomic structure of Sn(PbO$_2$)$_2$~\cite{Gavarri81p81} allows for the possibility of utilizing material design to shift the Fermi energy closer to the double Dirac point. For example, when Sn is substituted with In or Sb, which have one less or one more valence electron than Sn, respectively, the electron filling can change from zero to four (mod 8), possibly hosting the double Dirac point close to the Fermi energy. 

\subsubsection{Pb$_3$O$_4$ in space group 135}

\begin{figure}[h]
\includegraphics[width=0.45\textwidth]{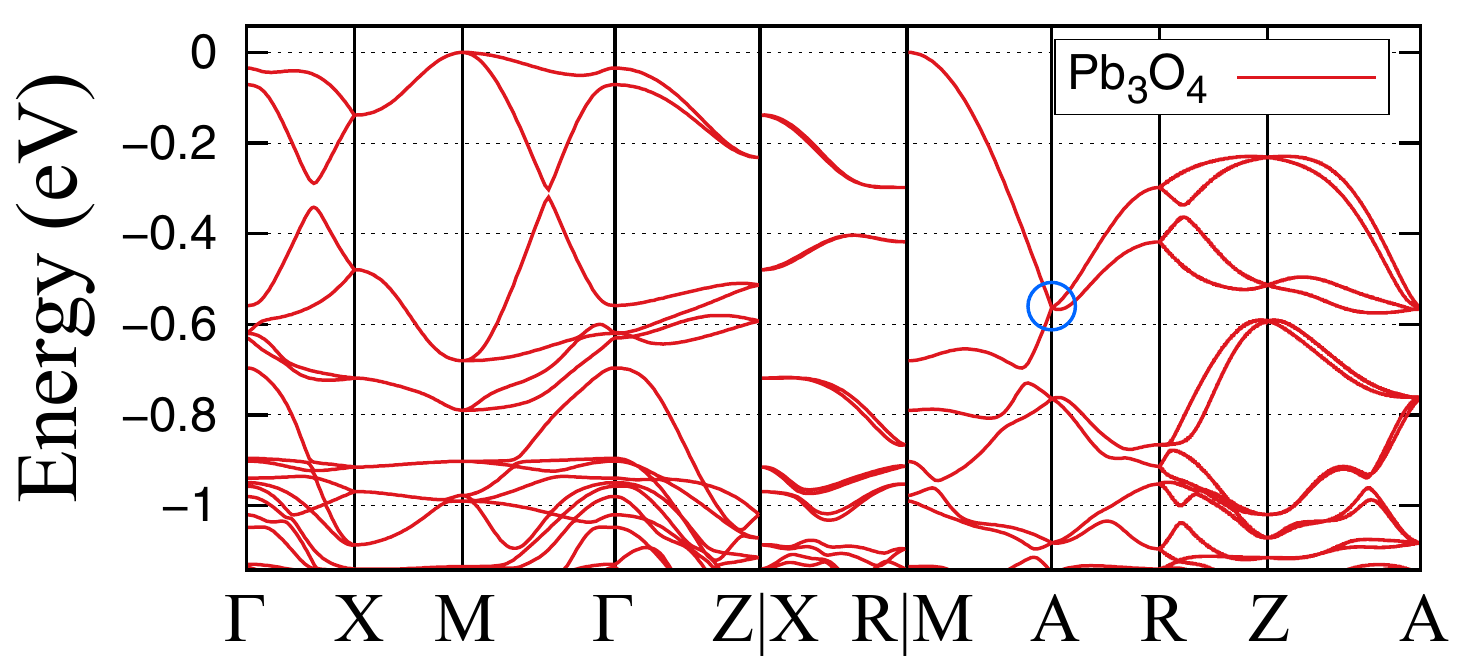}
\caption{\label{fig3}
Electronic band structure of Pb$_3$O$_4$ in space group 135. 
}
\end{figure}

Figure\ \ref{fig3} shows the band structure of Pb$_3$O$_4$ in space group 135. Pb$_3$O$_4$ is isostructural with Sn(PbO$_2$)$_2$ with Sn being substituted by Pb. The band structure also resembles that of Sn(PbO$_2$)$_2$. A double Dirac point resides at $ A$ in the valence band region near $E =$  -0.57 eV.                                       

\subsubsection{GaMo$_3$ in space group 223}

\begin{figure}[h]
\includegraphics[width=0.45\textwidth]{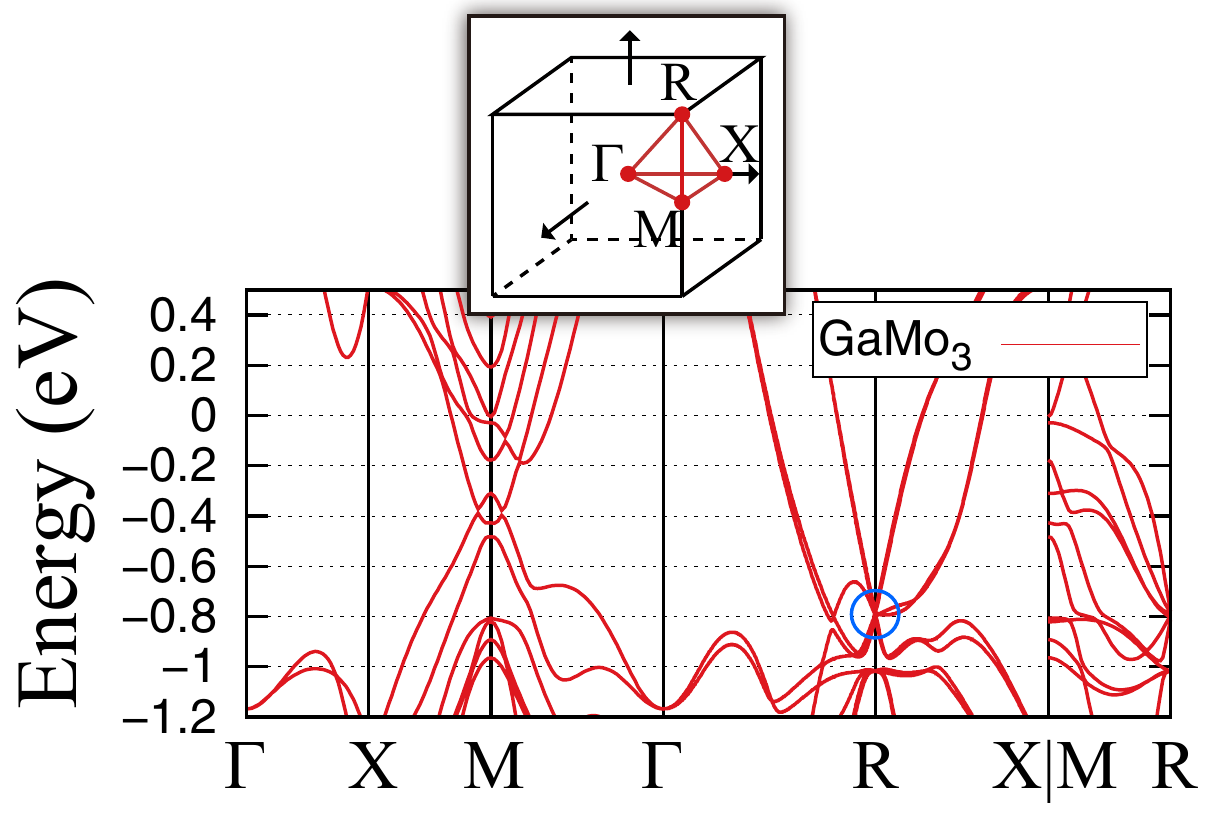}
\caption{\label{fig4}
Electronic band structure of GaMo$_3$ in space group 223. 
 }
\end{figure}

The electronic energy bands of GaMo$_3$~\cite{Bornard73p205} are drawn in Fig.\ \ref{fig4} along the high--symmetry lines of the cubic BZ, depicted in the inset. The position of the double Dirac point is indicated by a blue-colored circle.

\end{document}